# Software-Defined Network for End-to-end Networked Science at the Exascale

Inder Monga, Chin Guok, John MacAuley, Alex Sim
Energy Sciences Network, Lawrence Berkeley National Lab, Berkeley, CA, USA
{imonga@es.net, chin@es.net, macauley@es.net, asim@lbl.gov}

Harvey Newman, Justas Balcas
Division of Physics, Mathematics and Astronomy, Caltech, Pasadena, CA, USA
{newman@hep.caltech.edu, jbalcas@caltech.edu}

Phil DeMar
Computing Division, Fermi National Accelerator Laboratory, Batavia, Illinois, USA, demar@fnal.gov

Linda Winkler
Computing, Environment and Life Science Division, Argonne National Lab, Argonne, Illinois USA
winkler@mcs.anl.gov

Tom Lehman
Virnao, Arlington, VA USA, tlehman@virnao.com

Xi Yang
Mid-Atlantic Crossroads, University of Maryland, College Park, MD USA, maxyang@umd.edu

*Abstract*—**Domain science applications and workflow processes are currently forced to view the network as an opaque infrastructure into which they inject data and hope that it emerges at the destination with an acceptable Quality of Experience. There is little ability for applications to interact with the network to exchange information, negotiate performance parameters, discover expected performance metrics, or receive status/troubleshooting information in real time. The work presented here is motivated by a vision for a new smart network and smart application ecosystem that will provide a more deterministic and interactive environment for domain science workflows. The Software-Defined Network for End-to-end Networked Science at Exascale (SENSE) system includes a model-based architecture, implementation, and deployment which enables automated end-to-end network service instantiation across administrative domains. An intent based interface allows applications to express their high-level service requirements, an intelligent orchestrator and resource control systems allow for custom tailoring of scalability and real-time responsiveness based on individual application and infrastructure operator requirements. This allows the science applications to manage the network as a first-class schedulable resource as is the current practice for instruments, compute, and storage systems. Deployment and experiments on production networks and testbeds have validated SENSE functions and performance. Emulation based testing verified the scalability needed to support research and education infrastructures. Key contributions of this work include an architecture definition, reference implementation, and deployment. This provides the basis for further innovation of smart network services to accelerate scientific discovery in the era of big data, cloud computing, machine learning and artificial intelligence.**

*Keywords—Intent based networking, end-to-end orchestration, intelligent network services, distributed infrastructure, resource modeling, software defined networking, real-time, interactive*

# 1 Introduction

Networked systems are evolving at a rapid pace toward programmatic control, driven in large part by the application of software to networking concepts and technologies, and evolution of the network as a critical subsystem in global scale systems. This is of interest to major science collaborations that incorporate large scale distributed computing and storage subsystems. This software-network innovation cycle is important as it includes a vision and promise for improved automated control, configuration, and operation of such systems, in contrast to the labor-intensive network deployments of today. However, even the most optimistic projections of software adoption and deployment do not put networks on a path that would make them behave as a truly smart or intelligent system from the application or user perspective, nor one capable of interfacing effectively with facilities supporting highly automated data analysis workflows at sites distributed around the world.

Today, domain science applications and workflow processes are forced to view the network as an opaque infrastructure into which they inject data and hope that it emerges at the destination with an acceptable Quality of Experience. There is little ability for applications to interact with the network to exchange information, negotiate performance parameters, discover expected performance metrics, or receive status/troubleshooting information in real time. As a result, domain science applications often suffer highly variable (from excellent to poor) performance, especially so in highly distributed data intensive environments.

Indeed, the ability to interact and negotiate with the network infrastructures within a science ecosystem should be a hallmark of truly smart networks and applications. The current static, non-interactive network infrastructures do not have a path forward to assist or accelerate domain science application innovations. We therefore envision a new smart network and smart application ecosystem that will solve these issues and enable future innovations across many Research and Education (R&E) domain science communities. The Software-Defined Network (SDN) for End-to-end Networked Science at the Exascale (SENSE) [1] project has developed an architecture and implementation to address this vision. Key contributions of this work include an architecture definition, reference implementation, and deployment. This provides the basis for further innovation of smart network services to accelerate scientific discovery in the era of big data, cloud computing, machine learning and artificial intelligence.

The SENSE solution was built upon the previous Software-Defined Networking (SDN) work [2], which has been a subject of much discussion and research over the past decade. The crux of the SDN concept is software control and programmability of network elements and resources in a manner which enhances network services, management, and resource use. Multiple frameworks and systems have been developed which carry out the basic goal of software-controlled services across a heterogenous mix of network elements. While many of these systems are open source, the reality is that significant deployments in the field have been few and limited. These deployments are typically either small isolated systems, or vertically integrated systems from larger operators who have complete control over all the network resources needed by the higher-level applications being served. A lesson learned is that basic software-defined control functionality does not solve many of the key issues as needed to enable pervasive deployment of end-to-end automated services across general cyberinfrastructure. These issues include handling of multiple administrative/control domains, resource state hiding/visibility, scalability, and real-time responsiveness, all of which need to be tailorable for specific deployments and application requirements.

Initial SDN work was mostly focused on the SDN controller south bound interface and network element control mechanisms. Multiple mechanisms utilizing technologies such as OpenFlow [3], NETCONF and Yang [4], and others were defined. It was eventually realized that the SDN controller northbound interface was where the users obtain services, and that the exact mechanisms used on the southbound side were not as interesting from a user/application perspective. Subsequent SDN controller northbound work resulted in several systems focused on specific use cases and point solutions.

Partly because of these issues, much of the SDN research and development energy has transitioned to orchestration services/systems. There are now multiple open source orchestration projects [5][6][7][8], which include mechanisms to interact with multiple underlying SDN systems via their SDN Controller

Northbound Application Programming Interface (API). However, there has been relatively little architectural work to define what is needed in an SDN controller northbound interface to enable orchestration systems to address issues associated with systems that extend "east and west" across multiple administrative/control domains, heterogeneity among the architectures and policies in each domain, as well as resource state hiding/visibility, scalability, and real-time responsiveness. In addition, there has been even less work done in building systems which will enable the desired smart-application-to-smart-network interactive ecosystem, where in the case of major science programs the "application" may itself be a data management system that deals in real-time with computational workflow among sites on several continents.

In summary, current SDN and orchestration technologies have the following issues which inhibit development of an integrated, interactive smart network and smart application ecosystem:

- Current SDN Technologies - The SDN Controller Northbound API solutions are narrowly focused designs which are typically driven by an underlying southbound API feature set. The opposite approach should be used, where the user-facing API should be developed based on user/application requirements, with the southbound API and feature sets correspondingly adapted. With such a layer-based orchestration architecture, the SDN Controller Northbound API should be constructed with the orchestration layer as its user.

- Current Orchestration Technologies - The current orchestration architecture and associated implementation projects have not defined the requirements and features needed for the API between the orchestrator and underlying SDN Controllers. This would be the SDN Controller Northbound Interface, which would also be the orchestrator southbound interface. These requirements and the orchestrator functionality should be driven by the user/application requirements and therefore reflected in the orchestrator northbound interface.

- Combined SDN and Orchestration Technologies - The current solutions are focused on traditional service provisioning, customer onboarding, and operations/maintenance. While updated technologies such as Network Functions Virtualization (NFV) and automated provisioning are being employed in service of this paradigm, the service and use model is not much changed from a customer perspective, outside of the ability to initiate automated functions. These systems are not currently on a path to provide the degree of realtimeness and interaction needed for the smart network and smart application ecosystem envisioned by the Research and Education (R&E) community [9]. For example, the information exchange between the SDN Controller and orchestrator is not designed with the ability to i) include/exclude real-time states, ii) adjust the degree of resource/topology sharing/hiding as required by local policy and/or user requirements, and iii) tailor operations to optimize scalability or real-time responsiveness. In addition, the orchestrator functions are not designed to take advantage of enhanced interactions with underlying SDN controllers with a focus on interaction, real-time responsiveness, and intelligent services.

- End-to-End Solutions - Current SDN and Orchestration solutions are not end-to-end in the context of application workflows. Domain science application workflows need solutions which manage all resources along the end-to-end path. This needs to include the networking stack inside the end systems, as well as the devices along the network path.

The problem statement and solution objective which motivates this work are as follows:

**Problem Statement:** Current SDN and orchestration systems do not supply the degree of interaction, realtimeness, and intelligent network services needed for the next generation of domain science workflows. An integrated smart network and smart application ecosystem is needed to enable application workflows to ask questions, iterate on solutions, receive recommendations, and access full life-cycle status and troubleshooting information. Future SDN enabled infrastructures need mechanisms to provide topology and state information in real time based on fine-grained policy, scalability, and service objectives. End system resource management needs to be integrated into the orchestration of end-to-end network resources. In the longer term end-to-end system operation needs to be monitored at several layers with enough granularity, to supply a foundation for future system optimization through such mechanisms as reinforcement learning.

**Solution Objective:** Domain science application workflows need *real-time*, *interactive*, *end-to-end* orchestrated SDN services across large, distributed, multi-domain networks.

In this paper we present the SENSE model-based orchestration system which operates between the SDN layer controlling the individual network regions, and users/applications needing a variety of end-to-end network services. The SENSE system provides a solution to the identified problem and includes a novel set of APIs and methods for interactions with users/applications, as well as with the underlying software-controlled network infrastructure.

Multiple science community vision and requirement reports [39][40] have identified these types of network services as being important for the next generation workflows, including many that will be driven by Exascale computing resources and big data. Also driving the need for these types of network services is the emerging DOE Superfacility concept which includes the seamless integration of multiple, complementary DOE Office of Science user facilities into a virtual facility to fundamentally transform and accelerate the scientific discovery workflow. The SENSE system provides the mechanisms needed to synchronize and coordinate the connection of multiple distributed compute, storage, and instrument resources with deterministic performance. These intelligent interactive services provide methods for application driven workflow planning and operations assistance, which will be needed to realize the Superfacility vision.

As an example, the current Exascale for Free Electron Lasers (ExaFEL) [15] workflows utilizing multiple data transfers over best effort network paths are being replaced by SENSE services providing deterministic network paths capable of supporting real-time data streaming directly to compute memory or burst-buffers. This mode of operation will also support computational steering, where instruments use data streaming to drive preliminary compute results which are then used to calibrate and guide experiment configurations to create real-time science feedback loops.

It should also be noted that this new class of smart, interactive networked services is not expected to replace the existing best effort routed IP services in use today. Most science data flows will continue to use traditional IP routed services. However, based on historic use patterns and formal requirement studies [41], there will be a set of science-driven use cases which do require these types of advanced, end-to-end network services. While these use cases will make up a small subset of the total science data flows, they are expected to be responsible for the majority of the bandwidth utilization. Another important observation is that the traditional IP routed services and these advanced smart network services will need to run over common infrastructure, as a key aim is to not require separate or parallel network infrastructures. Advanced smart network services can be realized via using advanced IP features, such as segment routing, or direct access to the underlying Layer 2 or Layer 1 infrastructures over which traditional IP network services are run.

The remainder of this paper will describe the SENSE Solution (Section 2), SENSE Services Implementation (Section 3), Testbed Deployment (Section 4), Use Cases (Section 5), Performance Evaluation Analysis, Results and Analysis (Section 6), and Summary and Future Plans (Section 7).

## 2 SENSE Solution

As summarized in the introduction, there are several key features which are missing from current solutions as they relate to domain science research and associated cyberinfrastructure systems. SENSE enables a new application to networked system interaction paradigm, which includes the following capabilities in response to the problem statement and solution objective:

- <u>Intent Based</u> - The ability for an application to submit a service request in the form of a high-level statement of desired results or outcomes, as opposed to a specific set of network centric inputs. The format of an Intent based interface will be customized based on application specific requirements. In some situations, an intent may be expressed based on a highly abstracted network view with performance metric annotations. In other situations, the intent form will be expressed in the context of application specific resources, end points, and references. The SENSE system is designed to apply a DevOps (Development Operations) model to the interface construction, which is enabled by a rich semantic model-based infrastructure description which allows for variable levels of abstractions and infrastructure/services relationship tracking.

- Interactive - The ability for an application workflow agent to engage in a "conversation" via a bi-directional exchange with the network as part of workflow planning. This conversation can include discovery of available services, asking "what is possible" or "what do you recommend" types of questions, engaging in iterative negotiations prior to actual service requests, or full-service life-cycle status and troubleshooting queries. This can be extended to processes that drive adjustments or remedial actions to maintain system performance and/or task progress, and to balance among competing demands on the available resources.

- Realtime - This term has many different meanings and time scales depending on the situational context. For the purposes of this project, the problem space is large scale multi-domain, orchestrated SDN services. Each of the full lifecycle activities of resource discovery, provisioning, service status, troubleshooting, and feedback response loops may have their own requirements as it relates to real-time operations. In these contexts, realtime typically means a time scale of seconds to minutes. For example, provisioning an end-to-end path which consists of two Department of Energy (DOE) Laboratories High Performance Computing (HPC) facilities connected across a single wide area network, may have a response time in the 10s of seconds. A more complex end-to-end path with ten or more separate administrative domains, may have a response time of several minutes. A key objective of the SENSE design is to provide the mechanism where a tradeoff between realtimeness and scalability can be made at runtime by dynamic configuration. The SENSE model-based interface between the orchestration and SDN layer is designed to allow this tradeoff via controls that dynamically vary the real-time states which are included as part of the topology distribution. In addition, there are mechanisms which allow for on-demand discovery of real-time information and associated service parameter negotiation.

- End-to-End - The SENSE notion of end-to-end orchestrated SDN includes the multi-domain wide area, regional, and end-site networks as well as the network stack inside the end systems. The inclusion of the end-system networking stack is important from deterministic and automated service provisioning, monitoring, and troubleshooting perspective. The practical application of this approach is to manage the networking stack all the way to the network socket of the host operating system, virtual machine, or container where the application process is interacting with the network. This is designed to provide a foundation for applications such as science workflow management systems that coordinate the use of computational, storage and network resources.

- Full-Service Lifecycle Interactions - To optimize performance and adjust to changing conditions, applications need mechanisms to discover status and states during the service provisioning as well as during the service operational phase. This includes functions such as resource discovery, provisioning, service status, troubleshooting, and feedback response loops. The SENSE vision includes a continuous conversation between application and network for the full-service duration to enable new levels of application situational awareness.

The SENSE approach to end-to-end at-scale networking is based on software programmability and intelligent service orchestration. The SENSE orchestration architecture provides many performance and assurance benefits through application oriented services. These are enabled by some novel technologies, including a) hierarchical service-resource architecture (Section 2.2 for more details), b) unified network and end-site resource modeling and computation (Section 2.2.2 for more details), c) model based real-time control (Section 2.3 for more details), d) application driven orchestration workflow (Sections 2.4 and 2.5 for more details), and e) end-to-end network data collection and analytics integration (Section 2.6 for more details).

## 2.1 SENSE Key Functions

There are four main functions of the SENSE system:

- <u>SENSE Orchestrator North Bound Interface</u> – This is a highly customizable interface for application workflow agents to query regarding possible actions, recommendation, and/or request specific service instantiation. While a standard northbound interface has been defined, this interface is designed to be easily and rapidly customized for individual user requirements. The SENSE system has much data and intelligence regarding the underlying networked systems. This information can be customized for user consumption in a highly detailed or abstract manner.

- <u>SENSE Orchestration</u> - This includes the integration of resource model-based descriptions from underlying network infrastructures, the computation services to process resource models for user request responses, and the coordination of provisioning actions.

- <u>SENSE Orchestrator South Bound Interface</u> – This provides for a continuous exchange of topology descriptions which include an ability for the resource owners to tailor the level of abstraction and real-time states in accordance with local policies and service objectives. This is one of the key innovations of the SENSE system and is based on semantic web based graph models which provides a high degree of service flexibility and infrastructure owner controlled customizations.

- <u>SDN Layer</u> - The SENSE architecture relies on an underlying SDN layer; however, it does not require a specific SDN controller or system implementation. The SENSE architecture accepts that there will be a variety of deployed SDN solutions which will cover different network and administrative regions. SENSE provides mechanisms and functions to leverage these systems and guidance for how they can be fully integrated into orchestrated system. This typically requires existing SDN systems to implement the SENSE Orchestrator Southbound Interface as their controller Northbound Interface. Existing systems may accomplish this via native implementation of the SENSE API or via thin layer on top of their existing API which provides the proper interface. This technique of adopting underlying SDN systems for SENSE system integration has been used successfully as part of the SENSE system deployment on ESnet and other R&E infrastructures. Systems based on OpenDaylight (ODL) [10], Network Services Interface (NSI) [11], On-Demand Secure Circuits and Advance Reservation System (OSCARS) [12], and Open Network Operating System (ONOS) [13] have all been integrated into SENSE orchestrator operations. SENSE development and testing activities have demonstrated that valuable orchestrated services can be provided using these existing SDN systems as they are with no internal modifications. More advanced SENSE services can also be enabled by making some changes to these systems in the areas of topology description, abstraction, real-time states inclusion, and computation to support negotiations.

## 2.2 SENSE Architecture Components

Within the SENSE orchestration architecture, there are two distinct functional roles: Orchestrator and Resource Manager (RM). The interaction of Orchestrator(s) and RM(s) follows a hierarchical workflow structure whereby the Orchestrator accepts requests from users or user applications, determines the appropriate RMs to contact, and coordinates the end-to-end service request. The RMs are (administrative or technology) domain specific and are responsible for configuring and managing local resources.

An overview of the SENSE architecture is shown in Figure 1. At the lower layer are RMs covering various organizations from the R&E community who are part of the testbed deployment. These RMs create model descriptions for their infrastructure, in varying degrees of abstraction, and provide it to the Orchestrator. The SENSE Orchestrator absorbs and integrates these models to create an end-to-end model which provides a basis for subsequent intelligent infrastructure reasoning and service provisioning. The SENSE Orchestrator is also responsible for providing an interface facing the science users. To support a variety of use cases, the Orchestrator includes a pluggable Model Computation Elements (MCE) architecture, which enables flexible and rapid custom service construction. The Orchestrator operates between the automation layer controlling the individual networks/end-sites, and the science workflow agents/middleware layer. This figure also shows planned future work integrating external network monitoring and telemetry data sources into the model descriptions and services computation.

### 2.2.1 SENSE Orchestrator

The SENSE Orchestrator is expected to be closely associated with a domain science collaboration/application (such as LHC/CMS [14] and ExaFEL [15]) and processes "high-level" context sensitive intents to determine what resources are needed and coordinate the requests of "lower-level" (or sub) intents to the corresponding RMs. As such, the SENSE Orchestrator performs the following functions:

- Model Receipt - Receives model-based resource descriptions from multiple RMs.
- Intent Receipt - Receives and responds to the user's "high-level" intent requests (which is defined within the context of the user's domain science collaboration/application).
- Intent Processing - Renders the user's "high-level" descriptive intent request into "low-level" prescriptive requests for required resources.
- Resource Computation - Performs multi-constraint resource computation (based on AuthN (authentication) / AuthZ (authorization), resource availability, and other parameters) to determine the appropriate and necessary resources needed and which RMs to contact. The AuthN functions are utilized to create the trusted relationship between the SENSE Orchestrator and the individual SENSE-RMs. The SENSE orchestration layer AuthZ/AuthN functions are based on industry standards such as OpenID [42], OAuth [43], and InCommon [44].
- RM Workflow - Coordinate requests and replies from RMs and feedback the results to the user accordingly.
- Status Queries - Support queries by the user for status and states.
- Notifications - Provide resource notifications to the user as necessary.

The SENSE Orchestrator can take on different functionality, customized to the domain science needs based on the experiment, compute, storage, network, and other resources available. A SENSE Orchestrator North Bound Interface (SENSE Orchestrator NBI) is provided to accommodate such needs via service-oriented intent based interactions. This interface is discussed in Section 3 along with the intent design, negotiation mechanisms, and workflow operations.

### 2.2.2 SENSE Resource Manager

The SENSE Resource Manager (SENSE-RM) is tied to a domain with physical resources, such as a Wide Area Network (WAN), a Regional Network, or a Site (with Science DMZ [16] resources). The SENSE-RM is responsible for management of its domain-specific resources and includes the following functions:

- Model Generation - Provides (appropriately scoped and abstracted) model-based resource descriptions.
- Orchestrator Interactions - Receives and responds to the "low-level" intent requests from the Orchestrator.
- Resource Computation - Performs multi-constraint resource computation (based on authentication/authorization, resource availability, other parameters) to determine the local resources appropriate and necessary to service the request.
- Resource Provisioning - Coordinates resource allocations/commitments, provisioning, and de-provisioning with local controllers as necessary.
- Status Query - Supports queries by the SENSE Orchestrator for status and state.
- Resource Notification - Supplies resource notifications to the SENSE Orchestrator as necessary.

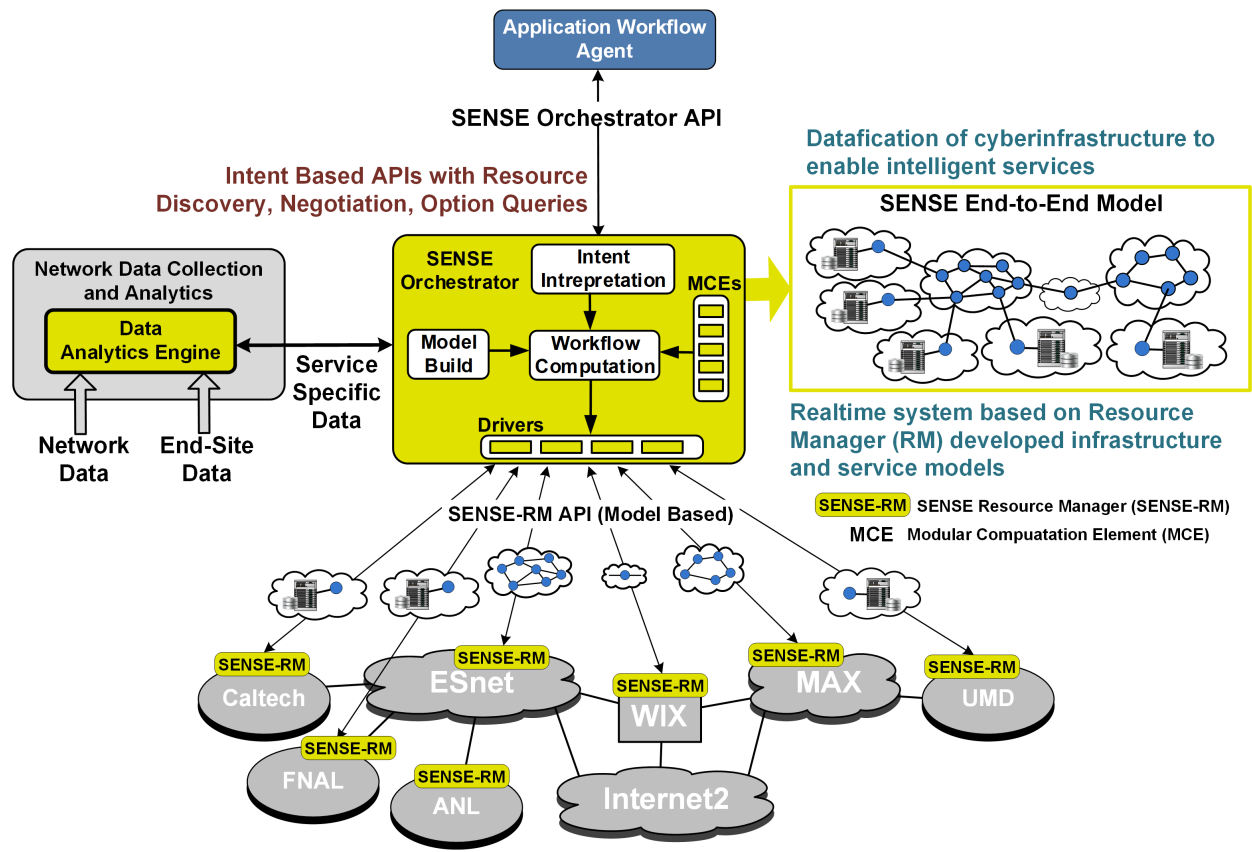

**Figure 1 SENSE Architecture**

The SENSE-RMs are specific to an administrative domain. However, within a single administrative domain, multiple instances of RMs may be deployed based on the distinct technology regions (such as Data Transfer Nodes (DTNs), optical packet/transport, Layer2, and/or OpenFlow resources). Conversely, a SENSE-RM may model multiple technology domains as a single resource description. For example, a network may have distinct switches and routers which provide layer 2 and layer 3 services correspondingly. However, the domain may instantiate a single RM which provides a unified resource description characterizing both sets of resources. The current SENSE implementation includes the following types of SENSE-RMs: Network Resource Manager (Network-RM) and Data Transfer Node Resource Manager (DTN-RN). There are multiple types of Network-RM which are tailored toward interoperation with specific types of underlying SDN systems. The DTN-RM is currently evolving into a Site Resource Manager (Site-RM) where it will manage local site networking in addition to the end host networking stacks.

### 2.2.3 Many-to-Many Relationship between SENSE Orchestrator and SENSE-RM

The SENSE Orchestrator should not be confused with a central orchestration service for all applications. Instead, multiple SENSE Orchestrator instances can independently serve different organizations, collaborations and application groups. The primary motivation for this architecture is that each scientific collaboration or workflow may have unique security, resource computation, and access policies/mechanisms. This allows each collaboration to implement fine-grained authentication, authorization, resource utilization, and security functions in accordance with its collaboration group policies. Each SENSE Orchestrator instance in turn has a unique trust relationship with multiple SENSE-RMs. This facilitates scalability in that an RM does not need to manage authentication, authorization, and policy information at the individual end-user level. The RM can enforce policies against the identity of the requesting SENSE Orchestrator instance and the negotiated service parameters.

In addition, different collaborations may have access to different resources within a SENSE-RM's domain. For instance, one collaboration may be restricted to a certain set of network links, whereas another collaboration may not have the same constraint. By having distinct SENSE Orchestrator instances per collaboration, a SENSE-RM may publish different resource descriptions based on Service Level Agreements (SLAs) that it has with the SENSE Orchestrator instance. The SENSE Orchestrator instance in turn may perform resource computation and allocation with priorities and constraints that are unique to the collaboration.

SENSE-RM's can also receive detailed information regarding the individual requesting user, which may be desired to apply finer grained policy and resource management policies. The SENSE Orchestrator supports multiple industry standard AuthN (authentication) / AuthZ (authorization) mechanisms, such as OpenID [42], OAuth [43], InCommon [44], Shibboleth [17], and Kerberos [18]. This facilitates the exchange of meaningful user information between multiple Orchestrator and RMs in the distributed system.

An important SENSE architectural premise is that from an Orchestrator's perspective, the RM is the owner and in control of its underlying resources. That is, the RM is the source of "ground truth" regarding the resource topology and states. The RM decides what set of resource descriptions to provide to an Orchestrator, and how to process (accept, modify, or reject) service requests. An Orchestrator's role is to simply gather resource descriptions from multiple RMs and facilitate the computation and service provisioning coordination across multiple Resource Managers. A key benefit of the SENSE approach is that the RM only needs to concern itself with the resources it owns and controls. A RM does not need to think about end-to-end services or resources in other administrative domains. This architecture will require careful optimizations and tuning in the two key dimensions of data consistency and conflict resolution.

Data consistency refers to the accuracy of the resource description information held by the Orchestrators. This information is provided by the RMs and may be incomplete or become inaccurate over time. A key benefit of the model-based exchange between the RM and Orchestrator is that the amount of real-time data and the update frequency can be optimized based on service objectives and scalability realities. An RM may provide a resource model, which includes only topology information that is relatively static. For instance, a resource topology model may just include the fact that an End Site has a specific number of Data Transfer Nodes (DTNs) which are connected at 100 Gigabit per second (Gbps), and that the site has 100 Gbps connections to two different wide area networks. It should be noted that changes to this "static" information are still automated, with periodic model updates sent as needed. At the other end of the resource model update spectrum, the RM could continuously update its states to reflect the current services provisioned and resource usage. For example, the resource model could include real-time information regarding services which are provisioned to the DTNs, including VLAN and bandwidth usage. This model could be updated every second, resulting in an Orchestrator with a much higher fidelity view of the current resources states. The SENSE system was designed to allow the tailoring of the amount of real-time data in the model exchange. Deployments which have small total number of RMs, and service objectives which include rapid provisioning, may want to provide frequent model updates with more real-time data included. Deployments which include a large number of RMs, and provisioning times which can be longer, may opt for model updates with less real-time information and less frequent updates. Both of these approaches will allow for successful multi-RM services provisioning due to the conflict resolution features which are described next.

Conflict resolution refers to the fact that an Orchestrator will likely be interacting with multiple RMs. Likewise, an RM may interact with multiple Orchestrators. The SENSE approach is to leverage the fact that the RM is in control of its resources and can always optimize based on its knowledge of real-time states. We could have taken the approach of also adding Orchestrator-to-Orchestrator coordination. However, we felt that this would add unnecessary complication to the system. Instead, we built into the Orchestrator to RM interface the notion of real-time negotiation and hold times. The negotiation features allow the RMs to inspect a request from an Orchestrator and suggest specific resource usage or configuration changes to optimize based in its unique knowledge of real-time states. The hold times provide a mechanism for the Orchestrator to receive a promise for specific resources being available for a small window of time, sufficient for it to coordinate with other RMs as part of a multi-RM service provisioning event. These features, with an intelligent Orchestrator possibly engaging in multiple negotiation rounds with some RMs as needed, are intended to result in a high percentage of service provisioning success, and resource utilization optimization. Testing and detailed data analytics in this area will be the subject of future work and papers. In addition, the need for Orchestrator to Orchestrator coordination may be reevaluated based on future testing and deployments.

There will be a tradeoff between optimization for data consistency versus contention resolution. Increasing the data consistency by including more real-time information in the model with more frequent updates, will allow an Orchestrator to make better initial decisions, along with less utilization of the negotiation and hold time features. However, the inclusion of too much real-time information may cause scalability or stability issues. If there is too little real-time state information available to the Orchestrators, it will have to rely on the negotiation and hold time features to obtain the real-time information at service provisioning time, which may increase service responsiveness from a user perspective.

Fairness and contention management are also important considerations that must be addressed at both the data plane and control plane levels. From the data plane perspective, SENSE includes some services with resource guarantees which allow these issues to be managed at the control plane level as part of service instantiation. For services which operate across shared resources, standard mechanisms for traffic monitoring can be utilized to identify unacceptable levels of utilization by specific users. Control plane interaction is the area where the SENSE system presents some new challenges in the fairness and contention management dimensions. A combination of monitoring, policy enforcement, and cost structures will be used to ensure that a user does not "game" the system to the detriment of other users. The first line of defense relies on the RMs using their freedom to flexibly manage their total resource pool for overall optimization based on the infrastructure owner policies. In this manner, the amount of resources which can be reserved or dedicated to a specific set of users can be tailored and controlled. The second line of defense is that there will be a "cost" associated with reserving or using specific resources. The specific form of this "cost" structure is future work, and may include monetary considerations, usage of allocation credits, or real-time monitoring to highlight and publicize when resources are not being used effectively. Thirdly, all users accessing the control plane will be authenticated which will enable historical usage profiling. This will allow the implementation of explicit priority and/or fair sharing policies and algorithms which manage the user interaction dynamics. The overall goal will be to ensure that if resources are made available via the SENSE intelligent interface, they are used effectively and efficiently.

## 2.3 Ontology-Based Resource Modeling

Orchestrating end-to-end SDN services over large network infrastructures must address two classic challenges: *control automation* and *distributed coordination*. Automation in any complex system requires formation of a control loop. In one direction, control operation results in state changes in the infrastructures. In the other, control feedback and/or telemetry is desired to provide additional state awareness back to the orchestration layer. Unified resource modeling can supply semantics in both directions. With a proper level of abstraction, the orchestration intelligence can learn dynamic resource and service states and create new services with reduced chance of conflict and better efficiency. The same modeling semantics can also serve to synchronize the orchestration intent to resource and service states in the underlying infrastructures and thus close the control loop. We call this a full-stack model driven approach, which is also a complete service-oriented approach. Applying this approach helps solve the other challenge in distributed coordination. When all resource owners use unified, extensible models to describe their resources, services, and states, we effectively create a thin-API to introduce universal programmability to all the parties. Each party can engage in free-form provider-consumer relationships for any As-a-Service transactions and thus decentralize the service integration, orchestration and instantiation processes.

The above vision led us to the search for a standards-based, composable, extensible and scalable semantic representation for Resource Modeling. We settled with using ontology-based modeling based on Semantic Web technologies [19]. Semantic Web, or Linked Data, is a suite of well-established standards by the World Wide Web Consortium (W3C) for web applications to describe and interconnect resources or data. Among the standards, the Resource Description Framework (RDF) [20] defines ways for "how" to exchange data, i.e. syntax, while the Web Ontology Language (OWL) [21] defines ways for "what" to exchange, i.e. semantics. The RDF/OWL combination provides a solution for defining ontologies which allow machines/software programs to understand and reason about the data.

Based on RDF/OWL we developed a Multi-Resource Markup Language (MRML) [22] as the ontology base for extensive types of resources and services in large information infrastructures. The modeling framework is based on extensions to the Network Markup Language (NML) [23] ontology developed by the Open Grid Forum (OGF) [24]. As part of a DOE Advanced Scientific Computing Research (ASCR) research project, RAINS [25], extensions to NML were defined to allow other resource types in addition to network elements/topologies to be described and modeled. The base NML standard and these extensions define the MRML, which is used as the ontology basis for resource modeling in the SENSE architecture.

## 2.4 SENSE Orchestrator To Resource Manager API

From the resource providers' perspective, the SENSE RM API provides the mechanism for real-time *ontology-based data integration* of distributed and diverse resource domains into the SENSE orchestration. The SENSE Orchestrator manipulates the provided topology model to achieve its target goal, computes and expresses a model "delta" between the original topology and the desired topology, and then proposes this resulting delta to the RMs.

### 2.4.1 MRML Resource Modeling

The SENSE-RM API is based on a resource model exchange and manipulation paradigm. The SENSE Orchestrator queries multiple RMs for a resource model which describes the infrastructure and services available for use. The resource model provided by each RM includes a description of its local network and other resources such as Data Transfer Nodes (DTN) [26], storage systems, instruments, and compute nodes. This model description includes a definition of the interconnects to external resources which allows the SENSE Orchestrator to build a model-based connected graph with all the RMs in its query space. This end-to-end model-based graph provides the basis for the SENSE Orchestrator to respond to user requests and construct workflows for service provisioning interactions with the proper RMs. In the SENSE Orchestrator, Modular Computation Elements (MCEs) provide the mechanisms to translate high level intent based user requests into specific workflow orchestration steps and resource requests to individual SENSE-RMs. Additional details regarding the MCE functions and usage for custom workflow computations is provided in Section 2.5.

### 2.4.2 Model Driven Real-time Resource Management

Each RM describes its topology and resources in the form of an MRML document with version management to track changes over time. This model document defines all the semantics for the SENSE Orchestrator API. Therefore, the API operations are radically reduced, down to two: *model pull* and *delta push*. The latter is divided into two methods, propagate and commit, to support a transactional Two-Phase Commit (2PC) push process. This simple set of API methods will not need to change when resource types or services are modified. Since all information is embedded within the model, only the model processing functions will need to be adjusted. In SENSE, we also emphasize another important performance metric: *realtimeness*. We will discuss what this means for end-to-end resource integration and service orchestration, below.

- *Pull* Model - The SENSE Orchestrator receives a model-based resource description from each of the RMs in the end-to-end SENSE ecosystem. The SENSE Orchestrator integrates models from multiple SENSE-RMs to generate a multi-domain resource description model. The individual SENSE-RMs utilize local policy to determine what information is provided with regard to resources, abstraction degree, and any other factors based on use cases associated with an individual

SENSE Orchestrator. On the current SENSE Testbed, the SENSE Orchestrator is customized to pull RM models every 30 seconds. The HTTP "If-Modified-Since" mechanism is used to reduce redundant data pull. SENSE-RMs will be responsible for adjusting the abstraction degree and resource update frequency to satisfy the "realtime" requirements posed by the SENSE Orchestrator. The SENSE-RM API also provides an optional *subscribe-notify* mechanism for the SENSE-RMs to push model changes to the SENSE Orchestrator before the *Pull* call, for speedier updates.

- *Propagate* Delta - The SENSE Orchestrator processes intent-based service requests from the SENSE Orchestrator API and generates a "model delta" which will be used to communicate a potential action/provision request to the SENSE-RM(s). The SENSE-RM is not expected to take any provisioning action based on the Propagate Delta method. In response to the Propagate Delta method, the SENSE-RM should inspect, verify, and confirm the request of suggest revisions. For example, a specific VLAN may be requested in the Propagate Delta method, while the SENSE-RM would prefer another VLAN. In this case the SENSE-RM should indicate the modified VLAN request in the response via modifying the provided "model delta". As the propagate call is composed entirely of data transactions, it can be executed quickly. Experiment results reported in Section 6 demonstrate that a Network-RM running on the production ESnet, which include a resource model with over 100 network elements, can execute a Propagate Delta around in under 11 seconds on average. A host based DTN-RM can execute a Propagate Delta in under one second on average. A negotiation procedure has been built into this phase such that multiple rounds of fast propagate and feedback transactions can be performed, to achieve an updated real-time result that may be different than the original "delta". This real-time negotiation and update is necessary as the SENSE Orchestrator and SENSE-RM are in a many-to-many loosely coupled relationship that does not always allow for a complete "real-time" synchronization of the resource state information.

- *Commit* Delta - The SENSE Orchestrator uses this method to ask the SENSE-RM to commit the changes negotiated as part of the Propagate Delta exchange(s). This is where the SENSE-RM is expected to provision resources. As this procedure is normally time-consuming, it is separated from the transactional propagate method. The SENSE-RM API commit is always asynchronous so that none of the SENSE Orchestrator calls to the SENSE-RMs are blocked for long time periods. Polling-based status queries are used to check the result of each asynchronous commit. Again, an optional *subscribe-notify* mechanism is supported for the SENSE-RMs to call back to the SENSE Orchestrator for real-time updates.

## 2.5 Intelligent Orchestration and Model Computation Framework

The core of SENSE Orchestrator is StackV [27], a general-purpose open-source orchestrator for networked multi-services. StackV is implemented based on the full-stack model driven intelligent orchestration approach. From the very top of the stack, applications communicate to the orchestrator with an abstract service intent. Intents including those specifically for SENSE take different forms, for the convenience of users. The SENSE Orchestrator NBI translates each service intent into a so-called "Service Model Description and Abstraction", which is a formal MRML model that consists of abstract resources annotated with service policy statements. The abstract model data are then fed to a dynamic compile procedure and compiled into a model-based computation workflow. A computation workflow consists of a variety of Model Computation Elements (MCE) as intelligent functions assembled into an execution tree. Each MCE uses system model data, service model data and policy data as input and accomplishes a specific function such as resource placement and connection computation. The output will be more detailed service model data, which could be used as input for another MCE. When the computation workflow finishes successfully, a System Model Delta will be created that provides detailed model statements about what needs to change in the underlying infrastructures governed by RMs to satisfy the intent.

The benefits of model-based computation include i) eliminating conversions between external interface and internal data structures, ii) leveraging standard tools for data query, navigation, transformation and reasoning, and, iii) maintaining consistent data semantics through all the computation modules. In this framework, MCE is the basic computation module. The input and output of an MCE are both model data based on the RDF/OWL, MRML and policy ontologies. Each MCE instance computes for a specific purpose and produces a compiled workflow of execution instructions. For an example, a Layer-2 VLAN Connection MCE absorbs the initial service abstraction model that specifies connection terminals, bandwidth and schedule parameters. It then creates model statements for end-to-end layer-2 connection across end sites and wide area networks. The result is an updated service abstraction model which is exported together with some intermediate policy data. The intermediate policy data has dynamically generated resource constraints and interdependencies that add to the context of next step computation actions. In this example, it suggests new VLAN interfaces as related to terminal ports and requests for data-plane IP addresses on such interfaces. Then a Layer-3 Address Assignment MCE uses this new service abstraction model and policy data (which is more detailed than the original one) as input to perform its own computation and add layer-3 modeling statements to the further updated service abstraction model. StackV has implemented sophisticated logic to concatenate MCEs and merge computation results. The basic idea of this technique is to use SPARQL [28] queries to "shape" the output of an upstream MCE into custom JSON format and use JSONPath [29] queries to extract information and "fit" to the input required by downstream MCEs. Success in finishing the computation workflow means StackV has resolved all model abstractions and policy annotations in the final product and has converted an application intent into a System Model Delta. This "delta" can be pushed down to the SENSE-RM API for instantiation. This modular model computation framework enables SENSE Orchestrator to perform in-situ intelligent computation when working with both real-time model data from SENSE-RMs and interactive intents from users.

In the context of this project, the terms "intelligence" and "smart" refer to several related SENSE architectural features and capabilities. To provide a flexible and customizable set of interactive, real-time, intent-based services across distributed autonomous SDN infrastructures, the SENSE system needs to do many things, which when taken together represents a certain level of intelligence. These activities include the absorption of information from the underlying dynamic SDN layer, computing multi-constraint solutions, and engaging in subsequent interactions and negotiations, both on the orchestrator southbound and northbound interfaces. Another context for intelligence is from the user services perspective. Here the user, via the SENSE Orchestrator northbound intent based API, can ask abstract and open-ended questions. As part of this, the user can engage in a "conversation" via a bi-directional exchange with the network as part of workflow planning. This conversation can include discovery of available services, asking "what is possible" or "what do you recommend" types of questions, engaging in iterative negotiations prior to actual service requests, or full-service life-cycle status and troubleshooting queries. This constitutes a certain level of intelligence from a user perspective and is discussed in more detail in Section 3 (SENSE Services Implementation). The third context for intelligence is based on the SENSE architecture definition and vision, which includes the incorporation of real-time telemetry data to feed the orchestration algorithms. The expectation is that this data will also be used to feed future machine learning systems, which will provide a mechanism for enhanced SENSE operations. This network telemetry integration, machine learning, and artificial intelligence work is part of future work, and represents the plan for SENSE movement toward more intelligence as a key part of future services. Additional information on this is provided in Section 2.6.

## 2.6 Network Data Collection and Analytics Integration

Topological model and resource states are the basis for the SENSE Orchestrator intelligent computation for orchestration services. In the current SENSE Testbed, ESnet and many DTN end sites have deployed various monitoring and data collection and archiving mechanisms. The planned SENSE analytics solution will consolidate these existing resources into a functional utility engine that has distributed data collection, archiving and access endpoints, but uses common API and data schema definitions.

The expectation is that further integration of real time and historical network data through an analytics engine can provide improved quality of experience for users, through better understanding of end-to-end network states and more precise prediction of traffic trends. The analytics-based feedback will also help users better understand network conditions and options and refine their service intent requests. An extended SENSE architecture includes integration with a data analytics engine that collects network data from end sites and transport networks, and provides analytics pre-processing and feedback to the SENSE Orchestrator. It will collect extensive telemetry data from various monitoring and active measurement sources that reflect network resource utilization and real-time states. This data collection and analytics capability is not yet in place and is anticipated as part of future work.

The Data Analytics Engine will be a component external to the SENSE Orchestrator. Following the suit of model driven API design, the interaction between the Data Analytics Engine and SENSE Orchestrator will be based on the same resource model used for the orchestration and resource management functions. With per-user and per-service ownerships being annotated upon collection, data contents and formats will be customized based on service orchestration needs. In addition, the analytics data will be integrated with the existing MRML model through abstraction, reference and annotation processing. New MCEs will also leverage the custom, pre-processed, MRML friendly data from the Analytics Engine to compute improved results for existing service intents and provide answers to more complex intent questions. This includes finer grained and more accurate answers to the "what is possible" or "what do you recommend" types of questions. In general, the objective is to utilize historical and real-time telemetry data to provide the user with estimates regarding end-to-end performance, and recommendations about when and how to use the network.

The Service Specific Data bridge across the Analytics Engine and the SENSE Orchestrator will form a closed control-feedback loop. The orchestration results will be monitored and measured and provided as feedback for fine tuning of future orchestration computation. On the other hand, the SENSE Orchestrator will also provide information to the Analytics Engine to help verify and instrument the data collection and analysis more efficiently. Including telemetry-based data analytics in the control-feedback loop will enhance SENSE realtimeness and interaction capabilities for end-to-end orchestration.

## 3  SENSE Services Implementation

The SENSE system has been developed to operate in "Development Operations (DevOps)" mode, where custom services can be rapidly developed in response to individual application requirements. The general system philosophy is that while not "every" service imaginable can be implemented, almost "any" service can be. This philosophy results in a system design that resource states and capabilities are sufficiently available to allow the construction of many different services. The user requirements will be utilized to form the basis of the actual services. For each of these services, the user can interact with SENSE in the following modes:

- Immediate Provision - If SENSE finds a resource path which satisfies the application request, provisioning starts at once (after routine confirmations from both sides).

- What is Possible? - In this mode, SENSE simply conducts a "Resource Computation" and provides the results back to the requestor. No provisioning action is taken without further explicit requests from the user.

- Negotiation - One or more rounds of Resource Computation requests with subsequent provisioning request by the application user if desired.

In the context of SENSE services, the "network" includes the switching and routing elements *and* the network stacks of the end systems, such as Data Transfer Nodes inside Science DMZ facilities. The data plane capabilities associated with these services are:

- Layer 2 point-to-point with Quality of Service (QoS)
- Layer 2 multi-point with QoS
- Layer 3 Virtual Private Network (VPN) and Flow QoS

Additional details regarding these (and other) services, the supporting system architecture, use case integration, and testing results are provided in the subsequent sections.

From the user application perspective the SENSE Orchestrator provides services via a programmable northbound interface, called the SENSE Orchestrator NBI. The SENSE Orchestrator supports modular intelligent computation and arbitrary orchestrated services. The SENSE Orchestrator NBI is a customizable intent based API with an emphasis on end-to-end network connection discovery, computation, and intelligent services to support science workflows.

### 3.1 SENSE Orchestrator End-to-End Service and Intent Based API

The SENSE Orchestrator NBI service is designed to be customized based on individual use case requirements. An example service is a "Multi-Path P2P VLAN" where a user requests a 10G connection with hard-capped bandwidth QoS between DTN sites at NERSC and Caltech. An alternative service type, "Multi-Point VLAN Bridge" could be used to request a VLAN connection of three and more terminals. A Layer 3 service allows the dynamic creation or attachment of end site resources to a specific VPN. The intent requests are captured in a simple JSON document and sent to the SENSE Orchestrator NBI for processing. A service request message format and key field information example is listed below:

<u>Service Request Message Format (example instance values in italics)</u>
    service_type: *Multi-Path P2P VLAN*
    service_alias: *sc18-p2p-b1*
    connections:
        name: *connection 1*
        terminals:
            uri: *urn:ogf:network:nersc.gov:2013:server+dtn11.nersc.gov*
                label: *any*
            uri: *urn:ogf:network:caltech.edu:2013:server+xfer-2.ultralight.org*
                label: *any*
        bandwidth:
            qos_class: *guaranteedCapped*
            capacity: *10*
            unit: *gbps*

<u>Key Message Field Information</u>

- <u>service_type</u> - This field indicates the type of service being requested. The example value is "Multi-Path P2P VLAN", which allows for multiple point-to-point connections to be computed and provisioned as a group.

- <u>service_alias</u> - This field indicates a service specific name for this service instance.

- <u>connections</u> - This section defines the specific connections requests. There may be multiple individual connections included in a single request. The advantage to including multiple connections in a single request is that they will be computed and optimized as a group with regard to satisfying the user requests and also using the network resources efficiently.

- <u>name</u> - This field supplies a service specific name for this connection instance.

- <u>terminals</u> - This block names one or more endpoint pairs for each connection.

- <u>uri</u> - A uniform resource indicator which defines the endpoints where service should terminate. The values correlate to information in the resource models.

- <u>label</u> - A list of any constraints or preferences for connection labels which may be VLANS or other network flow space element. The example value "any" indicates that the SENSE system may select a value based on available resources.

- <u>bandwidth</u> - This section defines the type of bandwidth desired for a specific connection.
- <u>qos_class</u> - An indication of the type of QoS desired with options as described in Section 3.3. Supported QoS classes include guaranteedCapped (no burst over the capped limit), softCapped (allowing for bursting over the cap when extra bandwidth is available) and bestEffort. The example value is "guaranteedCapped".
- <u>capacity</u> - This field indicates the amount of bandwidth requested.
- <u>units</u> - This field defines the units for the above capacity value.

SENSE has developed an intent schema to describe complex end-to-end network connectivity, QoS and scheduling requirements for the intent-based API. Internally, SENSE Orchestrator converts such intents into an ontology-based MRML model and converges them into a full-stack model computation, transaction and integration process that performs service instantiations and life-cycle operations.

## 3.2 Interactive Service Negotiation Workflow

The SENSE Orchestrator NBI also includes a set of messages which allow applications to interact with SENSE as part of its workflow planning. This includes SENSE messages for service and resource discovery, asking questions about options, and seeking recommendations. These are referred to as query and negotiation features. Additional description and examples for these types of interactions are provided below.

### 3.2.1 Service Request with Queries

A service request can optionally include a "query" block in order to ask questions without initiating actual provisioning events. Using the above service intent as an example, the below "query" blocks "ask" questions about end-to-end QoS capabilities. The query response from the SENSE Orchestrator then performs the regular service computation, to provide answers to the questions posed in the "queries". In the below example, the question is "What is the maximum bandwidth possible for the indicated connection?". The query response from the SENSE Orchestrator answers that it can allocate 10G of guaranteed and hard-capped bandwidth at this moment. The response also reports that this end-to-end path has a 100 Gbps bandwidth capability, based on the combined allocated and unallocated resources.

<u>Query Request Format (example instance values in italics)</u>
```
bandwidth:
    qos_class: guaranteedCapped
queries:
    ask: maximum-bandwidth
    options:
        name: connections1
```

<u>Query Response Format (example instance values in italics)</u>
```
bandwidth:
    qos_class: guaranteedCapped
    capacity: 10000
    units: mbps
queries:
    asked: maximum-bandwidth
    options:
        name: connections1
        bandwidth: 100000
        units: mbps
```

### 3.2.2 Service Request with Negotiation and Multi-Round Interactions

An application workflow agent may "negotiate" with the SENSE system by engaging in multiple rounds of query request/response exchanges. As part of this negotiation, the SENSE Orchestrator will revise the intent and post a newer version to the same service session identified by the service instance ID found in the reply from the initial service request call. Following the above example, the user knows the maximum bandwidth is 10 Gbps for the requested end-to-end connection. However, this only applies at the instant of the last reply, and this gives the user only a rough idea of the available network capacity. Then the user could negotiate for a feasible schedule in a sliding window that is bounded by the maximum and minimum allowed bandwidth, as illustrated in the request segments below. Here the user asks for a *time-bandwidth product* of 1 Terabyte to be transferred within next 2 days with acceptable bandwidth between 2 and 10 Gbps, the SENSE Orchestrator provided a feasible solution for a transfer, that it accomplished between 10:00:00 and 10:26:40 ET on September 1$^{st}$ 2018 at a fixed speed of 5 Gbps. Additional information regarding the time-bandwidth product is provided in Section 3.3.2.

<u>Negotiation using Query Request Format (example instance values in italics)</u>
```
queries:
    ask: time-bandwidth-product,
    options:
        name: connection 1
        tbp-mbytes: 1000000
        start-after: now
        end-before: +2d
        bandwidth-mbps <=: 10000
        bandwidth-mbps >=: 2000
```

<u>Negotiation using Query Response Format (example instance values in italics)</u>
```
queries:
    ask: time-bandwidth-product
    options
        name: connection 1
        bandwidth: 5000
        unit": "mbps",
        start":"2018-9-01T10:00:00.000-0400",
        end":"2018-9-01T10:26:40.000-0400"
```

### 3.2.3 Reserve and Commit Service

Negotiation can be performed for many rounds until the user is satisfied with the reply. Once the user has settled on the final intent, it could use the *Reserve* method to reserve the service, which corresponds to the reply sent by the SENSE Orchestrator in the last round of negotiation as the final intent. In a final step, the user calls the *Commit* method to actually allocate the resources. Compared to a "soft" *Reserve* that is mostly a database operation, the *Commit* call is "hard" operation, which can take considerable time for some resources. The SENSE Orchestrator NBI offers both synchronous and asynchronous methods to execute the commit call. The complete intent API document for SENSE Orchestrator NBI is published at [30], which includes other components such as the service termination and discovery methods.

### 3.3 End-to-End Quality of Experience and Intelligent Services

Through ontology-based resource modeling and intelligent orchestration, SENSE provides a powerful solution for end-to-end QoS across many network domains. Bandwidth QoS only represents one aspect of *Quality of Experience* for data transfer application users. Many users also want deterministic or predictable time schedules for data transfers. In addition, some users would like to ask open-ended questions, so they can optimize their workflow operations based on the network services and status.

Through implementation of these sophisticated capabilities, SENSE represents an innovative end-to-end SDN service paradigm. In this paradigm, the network control plane is an intelligent system that integrates and orchestrates arbitrary end-to-end services through ontology based real-time resource modeling and modular model computation. Users or client agents can then ask intelligent and complex "What is possible" questions via an intent based, interactive and negotiable service interface. In the SENSE project, we are building a reference implementation that is specific to the big science models, controlling primarily data transfer and network resources. Further developments of this implementation will continue to provide more sophisticated intents adapted to complex situations encountered in actual field operations, and possible optimized responses using machine learning.

To add more specificity to these ideas of smart network services, we present four examples of SENSE service capabilities that represent the use of query, negotiation, and question features to enhance the overall user Quality *of Experience*.

### 3.3.1 Immediate QoS Provisioning

This is the most basic feature where a user asks for a specific connection service with a specific QoS level. The supported QoS classes include *guaranteedCapped* (no burst over the capped limit), *softCapped* (allowing for bursting over the cap when extra bandwidth is available) and *bestEffort*. For users who are not sure how much bandwidth to ask for, or want to check availability before provisioning they may first query for the maximum available using the "maximum-bandwidth" query statement as shown in the earlier example. Once the intent negotiation is concluded, the service will be reserved across all domains, and then immediately provisioned once it is committed.

### 3.3.2 Time-Block-Maximum Bandwidth (TBMB)

With this feature the user would like to know the maximum bandwidth available for a specific time period. This is the same as the immediate provisioning service, but adds the time dimension to queries, and scheduling for provisioning. As an example, the user may ask for the same 10 Gbps connection as above without requiring immediate provisioning. Instead the request is to schedule a bandwidth service to start and end at specific times in the future, and to provide the maximum possible bandwidth that is continuously available during that block of time. The query "ask" segment would include a "total-block-maximum-bandwidth" value along with "start" and "end" values.

### 3.3.3 Bandwidth Sliding Window (BSW)

SENSE also implements a feature for end-to-end bandwidth scheduling based on the "sliding window" concept. As an example, a user may ask to schedule a service lasting for 4 hours that can be scheduled flexibly within the next 2 days. This particular intent is called a "bandwidth-sliding-window". The query statement is only slightly different than TBMB in that it includes a "start-after" and "end-before" fields allow the SENSE system to flexibly identify a time block within that window.

### 3.3.4 Time-Bandwidth Product (TBP)

Another SENSE service intent is based on the concept of the "Time-Bandwidth Product" (TBP). For instance, an 8-hour transfer at 10Gbps represents a data volume, or TBP, of 36000 gigabytes or 36 terabytes. Allowing users to query and negotiate bandwidth and schedule based on a given TBP is provided to assist bulk data transfer focused workflows, as TBP is a good estimate of the total amount of data to transfer. As an example, user "queries" may be formatted to find a schedule for the transfer an estimated 10,000 megabytes (10 gigabytes) of data within a 2 day time window after October 1st 2018 8:00ET. The user would like to check for the fastest possible transfer speed using a "use-highest-bandwidth = true" option. Alternatively, the user can ask for the least bandwidth (or widest schedule) using a "use-lowest-bandwidth = true" option, or a bandwidth-bounded schedule using both "bandwidth-mbps >=" and "bandwidth-mbps <=" options. The latter will return a feasible schedule that satisfies both the time-bandwidth-product and the bandwidth upper and lower bounds.

## 4 Testbed Deployment

The SENSE solution architecture aims to address the problem of real-time interactive end-to-end SDN orchestration, which includes a complex set of issues and features revolving around distributed resource management, real-time modeling, multi-domain data integration, end-to-end orchestration, and intelligent service interface and interaction. The method for SENSE solution architecture design is more empirical than quantitative. Reference implementations and testbed experiments are the primary means to validating the design. In addition, a real-world, at-scale SENSE testbed deployment helps us evaluate its technical applicability for a wide spectrum of use cases, scenarios and application workflows.

The SENSE architecture, models, and protocols define methods such that new implementations can include the most advanced levels of smart interactive networked services. However, a key part of the SENSE vision is to allow adaptation to and deployment on existing facility deployments interconnected by production networks. The same features that allow for adjustment and optimization of realtimeness vs. scalability, also allow existing SDN deployments to adapt their use of the SENSE functions in a manner which is compatible with their underlying network infrastructure. This allows early deployment of SENSE services for testing and use case development, and also provides guidance for future upgrades of network automation systems. This approach allowed the deployment of SENSE services which operate on top of ESnet, DOE laboratory, and university production and testbed infrastructures. The result is a SENSE testbed which allows for real world testing and the ability to provide services to use cases which include connections to their production resources.

The SENSE testbed deployment had to deal with multiple existing deployed SDN systems. The SENSE system provides the mechanisms and infrastructure to leverage these systems and provide guidance as to how they can be fully integrated into the SENSE system. This requires existing SDN systems to implement the SENSE Orchestrator Southbound Interface as their controller Northbound Interface. Existing systems may accomplish this via native implementation of the SENSE API or via a thin layer on top of their existing API which provides the proper interface.

This technique of adopting underlying SDN systems for SENSE system integration has been used as part of the SENSE system deployment on ESnet and other R&E infrastructures. Systems based on OpenDaylight (ODL) [10], Network Services Interface (NSI) [11], On-Demand Secure Circuits and Advance Reservation System (OSCARS) [12], and Open Network Operating System (ONOS) [13] have all be integrated into SENSE Orchestrator operations. The SENSE development and testing activities have demonstrated that valuable orchestrated SDN services can be provided using these existing SDN systems as is, with no internal modifications. The practical implication of this approach so far is that existing SDN system capabilities may limit the degree of realtimeness or interactivity that SENSE can provide to the user's application workflows.

However, we have also identified the needed changes to these systems in the areas of topology description, abstraction, real-time states inclusion, and computations to support negotiation that will allow for full provision of the more advanced SENSE services. There are also opportunities for native implementation of a SENSE based SDN system which further enhances the ability to increase the realtimeness and interactivity of the orchestrated services. This typically involves a tighter coupling between the SENSE defined resource model generation, real-time states tracking, and resource control mechanisms. This native implementation approach was utilized for the SDN layer at end sites, and higher performance was observed in the subsequent testing activities.

A SENSE testbed has been deployed which includes a mix of development and production resources. This testbed is being utilized to develop and test the SENSE software, as well as test with domain science use cases. As shown in Figure 2, this testbed is deployed at multiple DOE laboratory and university facility sites. For the allocation of network resources, the SENSE system interacts with production provisioning systems of ESnet and other networks. For the end-system resources, a mix of production and prototype DTNs are deployed. For the production DTNs limited access is provided, resulting in tailoring the set of SENSE based dynamic configurations to match local site policies. This approach to use a mix of production and research resources enables experience with various real-world site deployments and considerations.

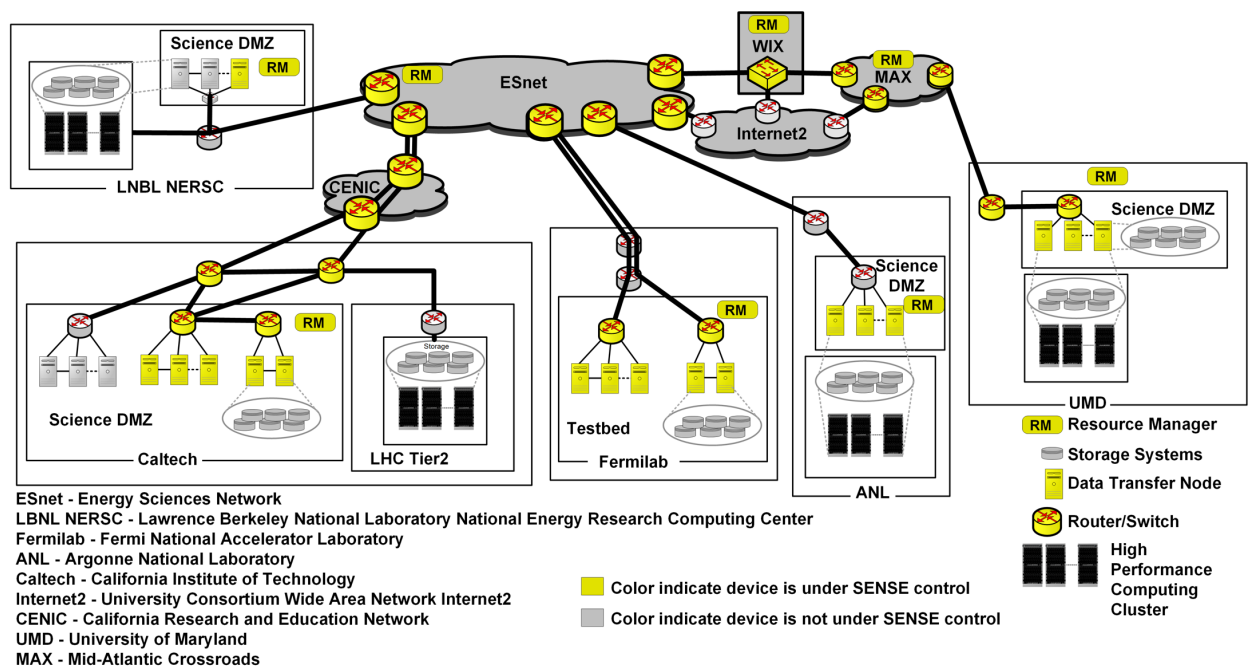

**Figure 2 SENSE Testbed Deployment**

The initial experimentation on the SENSE testbed were focused on validating the SENSE solution architecture design, evaluating soundness of intent based interactive service workflow for real-world use cases, and obtaining metrics on the key performance factors for realtimeness and scalability.

## 5 Use Cases

The SENSE project is now in a phase where use case integration is a key focus area. The main use cases currently under test are described below.

<u>Data Transfer Node Priority Flow</u>: Science DMZ located Data Transfer Nodes (DTNs) are a common method for moving data to/from compute facilities in the R&E community. For this use case, SENSE services are utilized to enable a "DTN Priority Flow Service". Since SENSE services are provisioned across the switching and routing elements *and* the network stacks of the end systems, this allows the creation of QoS-enabled paths that can be utilized for specific flows such that deterministic performance can be achieved regardless of the background traffic. The concept of operation is that these "SENSE enabled DTNs" can either be placed adjacent to current production DTNs as standalone transfer nodes, or SENSE software can be installed directly on the production DTNs. In either case, standard DTN operations and flows across the best effort routed IP paths continue as normal. When a SENSE flow is established between DTNs, this flow will receive priority access to network and host level resources. The best effort flows will continue, possibly at a reduced rate. This SENSE capability currently includes Layer 2 point-to-point,

Layer 2 multipoint, and Layer 3 VPN services. The workflow agent for this use case utilizes the "Time-Block-Maximum Bandwidth", the "Bandwidth-Sliding-Window", and the "Time-Bandwidth-Product (TBP)" SENSE features to instantiate Layer 2 paths with QoS. This workflow also demonstrates the "What is Possible?" and "Negotiation" feature sets. A description of SENSE services, as well as more information regarding testing for this use case is available here [31].

LHC/CMS Use Cases: The SENSE project is also working on use cases that integrate with Large Hadron Collider/ Compact Muon Solenoid (LHC/CMS) data movement and analysis workflows. SENSE integration with this science domain is focused in two areas:

- Rucio - This is a next-generation of Distributed Data Management system addressing high-energy physics experiment scaling requirements. Rucio was originally developed to meet the requirements of the high-energy physics experiment ATLAS and now is extended to support not only the LHC experiments but also other diverse scientific communities. Rucio uses File Transfer Service (FTS) to globally distribute the majority of the LHC data across the WLCG infrastructure. The SENSE team is implementing a plugin for FTS Service to be able to request network resources depending on the transfer queues and sizes between participating sites and do bulk transfer of files reliable from one site to another.
- nanoAOD - Another ongoing implementation is a new compact event form called the "nanoAOD" [34] that enables the rapid widespread distribution, ingest and real-time processing through a set of "PhysicsTools" of entire datasets of one to a few terabytes, that can be subsequently further analyzed on user's desktops and laptops.

The associated CMS analysis workflows and Distributed Data Management implementations currently under development, are planned to be accelerated and scaled up in terms of the number of simultaneous workflows supported, through the use of SENSE's interactive bandwidth allocation and management services, together with the RM services at a number of CMS sites, and high throughput data transfer applications such as FTS, XRootD and Caltech's open source Fast Data Transfer (FDT) [35].

Further related developments, underway through the NSF-funded SDN Assisted NDN for Data Intensive Experiments (SANDIE) project [36], include the use of Named Data Networking (NDN) and its caching and routing methods, to be supported as part of future SENSE services to expand NDN's ability to deal with larger scale data intensive workflows.

Exascale for Free Electron Lasers (ExaFEL): The objective of this use case is to stream nano crystallography diffraction data from SLAC National Accelerator Laboratory (SLAC) to National Energy Research Scientific Computing Center (NERSC) over the network in order to perform analysis on Cori, a Cray XC40, which has a peak performance of about 30 petaflops with 2,388 Intel Xeon "Haswell" processor nodes, 9,688 Intel Xeon Phi "Knight's Landing" nodes, and a 1.8 PB Cray Data Warp Burst Buffer. The feedback is provided afterwards to the beamlines in the form of 3D electron structure visualization. The workflow uses SENSE components to stream the data from the LCLS online cache at SLAC to NERSC compute nodes over network, and also orchestrates the SFX analysis processes to give near-real-time feedback to the experiment. For this use case, the ExaFEL application workflow agent utilizes the "Time-Block-Maximum Bandwidth" SENSE Service to provision the network path. This includes establishment of Layer 2 paths with QoS with time domain scheduling. More information about testing for this use case is available here [32].

BigData Express: BigData Express provides schedulable, predictable, and high-performance data transfer service for DOE's large-scale science computing facilities (LCF, NERSC, and US-LHC computing facilities, among others) and their collaborators. This project seeks to orchestrate the system, storage, and network resources involved in high-performance data transfers. From a network services perspective, BigData Express focuses on controlling local network resources supporting the end systems. For wide-area service, the BigData Express system utilizes SENSE services to provision paths across ESnet. The BigData Express workflow agent utilizes the "Time-Block-Maximum Bandwidth" and the "Bandwidth-Sliding-

Window" SENSE services to instantiate Layer 2 paths with QoS. This application also utilizes the "What is Possible?" and "Negotiation" features sets to co-schedule across multiple end-sites and network resources. More information about testing for this use case is available here [33].

## 6 Performance Evaluation, Results and Analysis

The SENSE solution architecture enables feature rich end-to-end services for many use cases. It is important to evaluate the design and implementation against our target problem in realistic settings and verify its performance in at-scale deployments. Results from testbed experiments will provide (a) reference performance metrics for integrating new network domains, end sites and facilities, and (b) ground truth in support of large-scale deployments and service operations. The performance study in this paper reflects an initial attempt to verify system functionality with a focus on real-time speeds and scalability metrics.

### 6.1 Performance Tests Setup

The overall performance evaluation setups include testbed configuration, experiments run, data collection and results analysis. For the baseline performance evaluation, we ran experiments and collected data from the actual SENSE Testbed that consisted of 8 select SENSE RMs, including 3 Network RMs for Production ESnet, ESnet Testbed and CENIC / PacificWave wide area domains, and 5 DTN-RMs for the NERSC/LBL, Argonne National Lab (ANL), Fermilab (FNAL), Caltech and University of Maryland (UMD) end sites. The SENSE Orchestrator and RMs were standalone software suites running on medium sized virtual servers (VM). Each VM typically had 4 vCPU cores and 8 GB of memory. The experimental testing plan was as follows.

- Compose - Compose a batch of 6 DTN-to-DTN service intents. Each represents a point-to-point (P2P) or multi-point (MP) layer-2 network connection service that requires transactions with 3, 4, 5, 6, 7 and 8 RMs respectively. These intents are described in Table I.
- Request - Request the batch of services to SENSE Orchestrator. Repeat the same batch when all the requested services have been orchestrated and go active.
- Collect - Collect performance data from SENSE Orchestrator and the 8 SENSE-RMs. This work mainly involves extracting beginning and ending timestamps for model generation, model pull, delta propagate, delta commit, model integration and orchestration computation events from logging outputs by all the participating systems.

**Table I. Service intents in one batch of testbed experiments.**

| Experiment Service Intent | # of RMs |
|---|---|
| P2P UMD - FNAL | 3 x RM |
| P2P NERSC - FNAL | 4 x RM |
| P2P NERSC - Caltech | 5 x RM |
| MP NERSC + Caltech + ANL | 6 x RM |
| MP NERSC + Caltech + ANL + FNAL | 7 x RM |
| MP NERSC + Caltech + ANL + FNAL + UMD | 8 x RM |

### 6.2 SENSE Testbed Baseline Experiment Results

Through the experiments, we collected data to verify the speed and scalability of the entire control-feedback loop for SENSE orchestration. We can break down the speed metrics into these three parts:

A. *Decision speed*: Service computation at the SENSE Orchestrator.
B. *Control speed*: Service model delta "propagate" and "commit" times from the SENSE Orchestrator to SENSE-RM.

C. *Feedback speed*: Model generation from SENSE-RM pull and integration times at the SENSE Orchestrator.

Data collected for (A) and (B) include per-service event times, which are presented in Figure 3. Item (C) is a per-system metric associated with the SENSE Orchestrator pull of a full model update from a SENSE-RM system. These results are presented in Figure 4.

There are three different types of SENSE-RMs referenced as part of these experiments: N-RM-OSCARS is a Python implementation of SENSE-RM on top of the latest ESnet OSCARS API, N-RM-NSI is a Java implementation wrapping around the Network Service Interface (NSI) that many R&E networks currently use for dynamic layer-2 circuit provisioning, and DTN-RM is a native implementation of SENSE model driven resource management for DTN centric end sites and local SDN implementations. The results are aggregated based on the system components involved in the orchestration: SENSE Orchestrator, N-RM-OSCARS, N-RM-NSI and DTN-RM. From Figure 3, we have the following observations.

1. Network RMs take a much longer time than DTN RMs to *propagate* a service delta. This is because both OSCARS and NSI have their own reservation system that adds overhead to the SENSE *propagate* process, while the DTN-RM can have this process built on native SENSE modeling without wrapping around a legacy system. The *propagate* is transactional, so *propagate* with RMs has to be synchronous, meaning the *propagate* times add up for all involved RMs in a service. The number of DTN end sites has very little impact as the DTN-RM *propagate* is very fast. The number of WAN domains the service traverses will have a major impact on the speed, as the average N-RM *propagate* operation takes about 11.2 seconds. The largest service in these experiments has 3 N-RM and 5 DTN-RM instances, bringing the total propagate transaction time to 48 seconds.
2. *commit* takes the longest times, because this is the process where resource allocation is actually executed. Unlike *propagate*, the *commit* process is asynchronous. Multiple commits are done in parallel, so that the total *commit* time equals the longest among the RMs involved in a service. 30 seconds of *commit* is quite common in these tests, although the maximum could be around 48 seconds when N-RM-OSCARS is involved. It should be noted that the actual RM commit could be very short. For example, a DTN-RM takes only a bit more than a second to finish end site configuration. The SENSE Orchestrator runs a periodic poll after the asynchronous *commit* call to identify the "finish" status which introduces an extra time penalty due to the fixed polling intervals. The SENSE-RMs can also use the optional *subscribe-notify* mechanism to call back to the SENSE Orchestrator for immediate update which eliminates this time penalty.

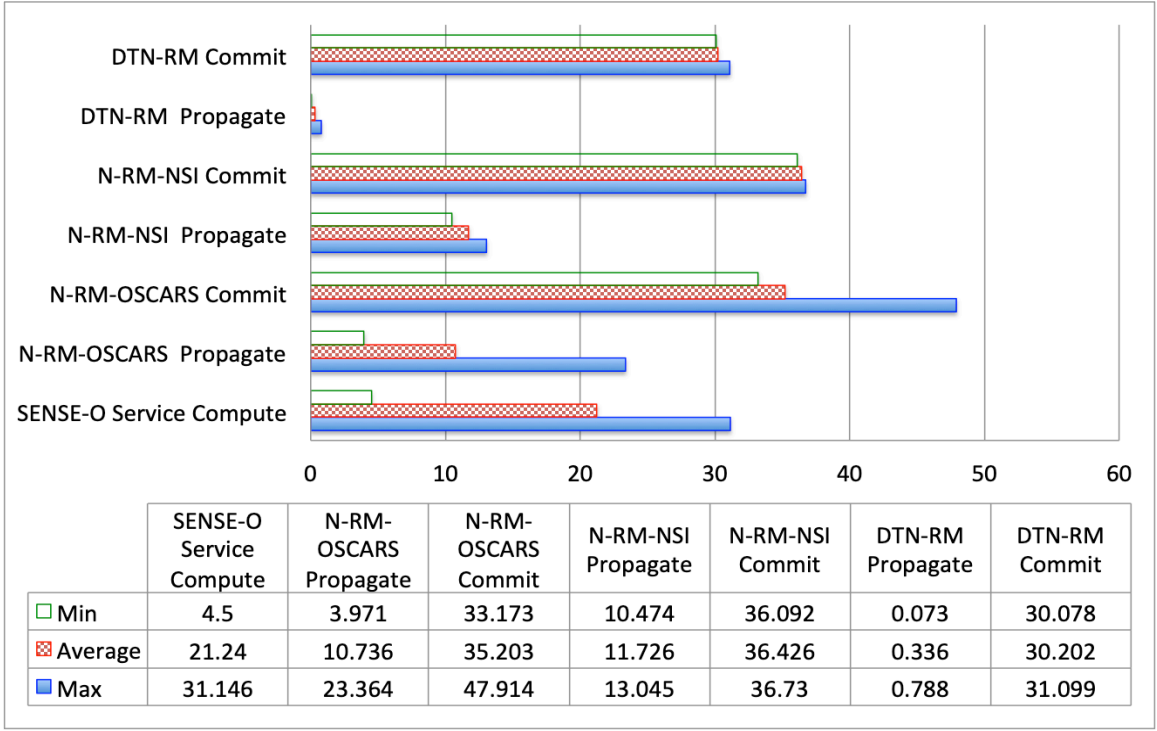

**Figure 3. SENSE Testbed service orchestration speeds breakdown (in seconds).**

3. The SENSE Orchestrator service computation times have the biggest deviation, ranging from 4.5 to 31.1 seconds. This is largely driven by the number of domains and provisioned paths in the requested end-to-end topology. A point-to-point (P2P) service spanning 3 or 4 RMs takes the least time for path finding, while a multipoint (MP) service spanning all 8 RMs takes the longest computation time.
4. Adding up the times for the various steps, we see that the full process of service control including computation, propagation and committing takes about 1 to 2 minutes at this scale. From the scalability perspective, the number of RMs involved in a service matters as it will add overhead to the service computation and delta propagation times.

The SENSE orchestration performance also depends on how quickly the SENSE-RMs can update their models, and how fast the SENSE Orchestrator can pull and integrate these model updates. The results shown in Figure 4 provide some insight into these factors.

1. The N-RMs need longer times than the DTN-RMs on average to generate a model, due to the overhead in communicating with the OSCARS or NSI API. However, this difference becomes less important when the network size (number of resources under the RM) increases. The N-RM-OSCARS takes a much longer time than the N-RM-NSI because the production ESnet controlled by N-RM-OSCARS is much bigger than the other two WANs controlled by the N-RM-NSI instances. Generating a model at the Caltech DTN-RM is also significantly longer than other DTN-RMs because Caltech has a more complex campus SDN component that adds to the model generation time for its DTN-RM.
2. Model pull times have the biggest range from milliseconds to many seconds, due to very different model sizes between RMs. The longest pull time of 24 seconds comes from the production ESnet N-RM-OSCARS, which generates a model from OSCARS on-the-fly with a *pull* API call. A planned improvement will separate the model generation procedure from the API method and reduce the pull time to sub-second. The SENSE-RM API recommends using the "If-Modified-Since" HTTP header, which allows the RM to return code 304 instead of the actual model content when no model update is available. SENSE-RM API also recommends HTTP content encoding with gzip compression. These options result in the shortest pull times for the two N-RM-NSI instances that have implemented them.
3. Model integration is a process in which the SENSE Orchestrator combines all the model pieces pulled from RMs into a connected *union* model. This requires CPU computation at the SENSE Orchestrator. Experiments show that this process is normally sub-second at this scale.
4. The SENSE Orchestrator model pulls from all RMs are asynchronous and done in parallel. The speed is determined by the combined model generation and model pull times at the slowest RM. The total feedback time, which is the system wide model learning time, can then be calculated by adding up the longest SENSE-RM model generation plus pull time and the SENSE Orchestrator model integration time. This total time is easily scalable as model integration time is the only limiting factor that will increase with the number of RMs in orchestration. The good news is that the model integration process is very fast.

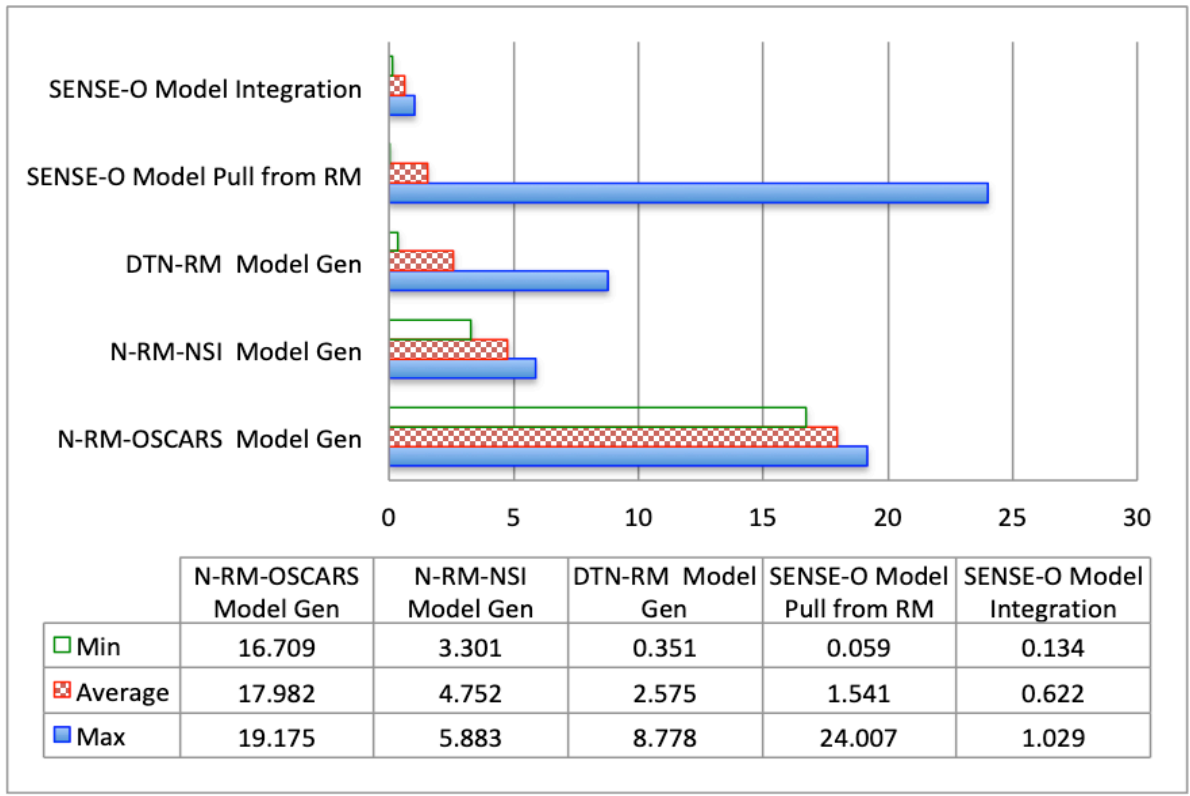

**Figure 4. SENSE Testbed system model generation pull and integration speeds (in seconds).**

To summarize, with a reasonable number and size of services on the eight-RM SENSE testbed, 1 to 2 minutes of decision making, and service control time is needed, and less than half a minute is needed for system wide model learning and update. We have identified and analyzed the factors that determine the overall decision and control time. The majority of these are related to the local computation and legacy API overhead, which can be improved through software internal tuning. At the SENSE-RM API level, several enhancements have been designed but not implemented. The results of our experiments have confirmed that these enhancements will potentially reduce the remaining overheads substantially. At the SDN infrastructure layer, we see that the choice of control plane technologies has a significant impact on performance, as manifested by the difference between pre-SENSE systems such as OSCARS and SENSE native implementations such as the DTN-RM SDN. The experimental results have shown that while SENSE can adapt legacy systems to work with SENSE-RM API, the control and feedback often becomes less than real-time. Enhancing such systems with SENSE native resource modeling, model update and negotiation mechanisms will greatly improve the overall performance and hence the Quality of Experience.

### 6.3 Scale-Out Emulation and Results

The baseline experiments summarized above provide the initial results and insights into real-time speed metrics. A bigger question now is: how would these findings fare in a much larger orchestration environment? To answer this question, we set up a SENSE emulation environment for scale-out experiments. We deployed a total of 25 emulated DTN-RMs on real-world end sites. We also created 42 NSI N-RM instances to emulate the global AutoGOLE network domains [37]. In this environment, both the DTN-RMs and N-RMs present realistic topological models to the SENSE Orchestrator. Model visualization of the 67-domain topology is shown in Figure 5. They interact with the SENSE Orchestrator through the same SENSE-RM API with realistic distributed transactions, except that the device-level resource allocation actions are emulated.

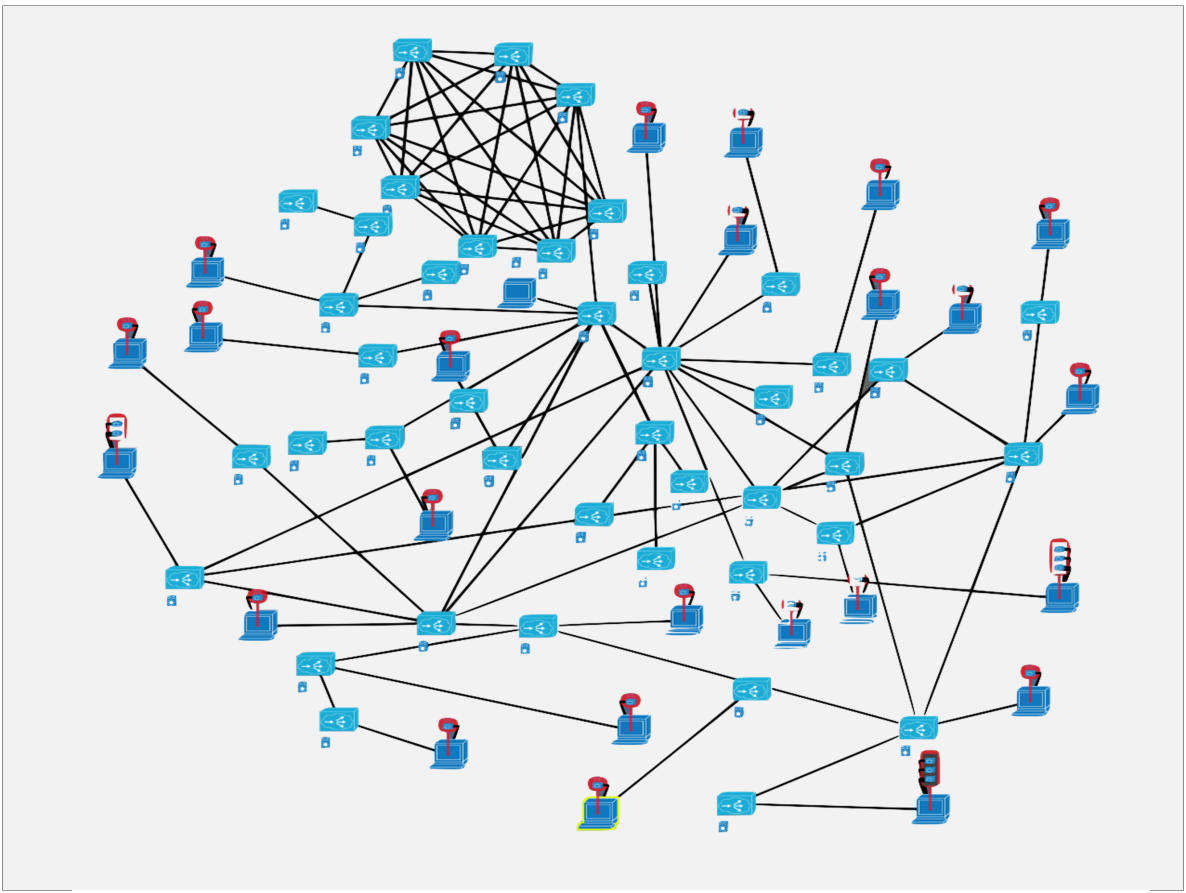

**Figure 5. SENSE-Orchestrator model visualization of the 67-domain topology (screen shot) for scale-out experiments.**

In the scale-out experiments, we generated multiple batches of service requests among select DTN end sites. Each batch consists of 20 service instances, requested simultaneously. When one batch is fully committed, a second batch will then be handled. This scale of SENSE deployment and service operations is very close to what is expected in real world global R&E infrastructures. We collected data from the SENSE Orchestrator and all 67 RMs and present the results in Figure 6.

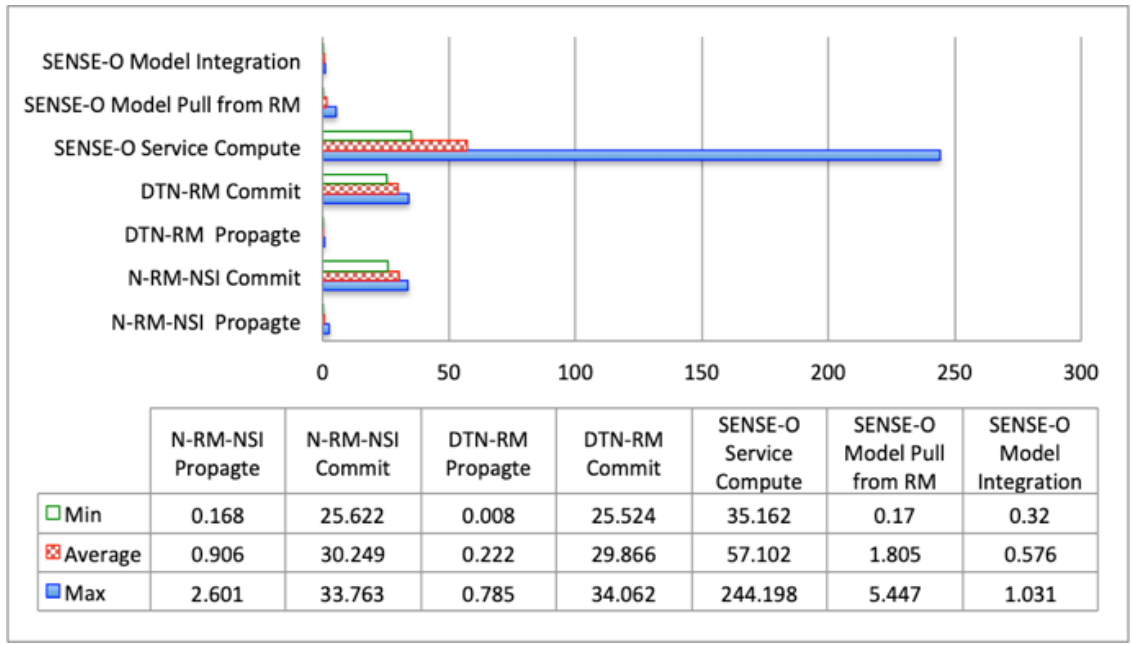

**Figure 6. SENSE scale-out experiments, with 42 WAN domains and 25 DTN end sites, results (speeds in seconds).**

These results confirm the projected scalability performance based on observations from the baseline experiments and supply further insights.

1. Model *propagate* and *commit* are per-RM events. They are not affected by the size of the testbed. Emulation makes them slightly faster than working with actual devices. Due to the SENSE Orchestrator poll based model verification, it takes around 30 seconds for a service to be committed, even though this is emulation. As the device-free commit takes no time, we could use the optional *subscribe-notify* mechanism to update SENSE Orchestrator for completion of commit operations. This should reduce the commit time to sub-second for emulation, and potentially a few seconds for real services.
2. SENSE Orchestrator scalability refers to its ability to handle large topologies (quantified by the number of independent resource models and elements) and high degrees of dynamism (many model updates due to tracking real-time states and/or high service provisioning frequency). This scalability is affected by three factors: model pull and integration from all RMs; per-service model computation; and per-service delta propagate and commit with involved RMs. The test results confirm the model pull and commit operations scale well with little performance impact due to their asynchronous and parallel performance mode of operation. The model integration and per-service propagate operations are synchronous and serial process, and there was some concern that they would not scale as well. However, based on the parameters of the testing described, there were no significant differences from the baseline experiments.
3. The biggest and practically only factor affecting the scalability is the service computation times at SENSE Orchestrator. We see an average of 57 seconds for each computation, and the maximum time can reach 4 minutes. One reason is that when dealing with a large network of 67 domains that consists of 92 nodes and over 200 links, the longest path could span over 9 domains and 11 nodes. Path finding takes significantly more time than for an 8-domain network. A more important reason is that with each batch of 20 service instances being requested simultaneously, the concurrent computation threads cause contention in a modest virtual server. SENSE Orchestrator is built with Enterprise Java and can be deployed in cluster mode with workloads balanced to multiple nodes. The current SENSE testbed has not taken advantage of this capability. A next generation of testbed will use a clustered SENSE Orchestrator deployment to eliminate the computation bottleneck.

From the scale-out experiments, we have proven that the overall SENSE architecture is optimized for parallel orchestration and is highly scalable. The orchestrator path finding algorithm and deployment size needs to be improved for large networks to reduce the per-service computation time. Combining that with the reduction introduced by RM-pushed commit notification, the whole SENSE orchestration will potentially only carry seconds of overhead. This will provide additional value for many real-time or time sensitive operations, as well as for major science programs with a global reach.

## 7 Summary and Future Plans

The SENSE system architecture and implementation presented utilizes model-driven datafication of cyberinfrastructure to enable intelligent network services. Science applications utilizing intent based APIs with automated resources discovery and negotiation enable a significantly different mode of operation as compared to current network usage modes. With the falling costs of 100 Gbps capable devices, powerful end systems are increasingly being placed at edge locations where high-bandwidth connections directly to regional and national networks will be the norm, and this trend will continue with the current emergence of 400 Gbps capable switches and the corresponding end systems. The Science DMZ based, National Research Platform Initiative [38] is an example of a high-performance end-system edge deployment. As a result, the expectation is that we are entering a cycle where network capacity will be easily overwhelmed by these advanced end-site and edge facilities. This indicates a need for methods to manage network resources and access them in a more intelligent manner, which includes providing the application agents with sufficient information so that they can plan and optimize their operations. The SENSE vision and solution is focused on these issues in anticipation of the time where unmanaged network utilization and extreme overprovisioning is no longer the preferred operational approach, or no longer feasible due to cost and/or technical considerations.

The SENSE project tackles a range of network research problems and has produced results that are interesting for many aspects of network research and practical operations. In this paper, we are focused on the centric problem of *real-time*, *interactive, end-to-end SDN orchestration*. We presented a SENSE Solution to this problem with detailed description of architectural functions, components and novel resource modeling, management, computation and orchestration mechanisms. We validated this solution as noted in the SENSE Services Implementation section and described how the SENSE Orchestrator supports sophisticated *interactions* for better quality of experience through an intent based interface and intelligent negotiation workflows. We further validated *realtimeness* and *end-to-end* aspects of the solution with description of the SENSE Testbed Deployment and Use Cases. In the Performance Evaluation section, we quantified the speed and scalability metrics through real-world experiments and emulations. This provided insights and assurance regarding the applicability of the SENSE solution to real-time, large, distributed, multi-domain environments. The experimental results presented in this paper also pointed us to several implementation issues that can be fixed or improved in a straightforward fashion via follow-up work. Initial prototyping [32] indicates that this solution provides a set of services which can greatly facilitate the realization of the emerging DOE Superfacility concept. This vision includes the seamless integration of multiple, complementary DOE Office of Science user facilities into a virtual facility to enable fundamentally greater capabilities. The SENSE system provides the mechanisms needed to synchronize and coordinate the connection of multiple distributed compute, storage, and instrument resources with deterministic performance and methods to assist in the application driven workflow planning/operations to realize the Superfacility vision. Key contributions of the SENSE work include the architecture definition, reference implementation, and deployment as the basis for further innovation of smart network services to accelerate scientific discovery in the era of big data, cloud computing, machine learning and artificial intelligence.

The SENSE solution architecture and services implementation creates many avenues for investigation and provides a platform to address interesting research questions. These issues revolve around the focus on interaction, negotiation, the degree of real-time state management and consideration at many levels of the decision and control operation process. Future plans include exploring some of these issues noted below as part of ongoing development and testing of the SENSE system:

- Real-time States - SENSE is designed to allow resource owner to flexibly adjust the amount and type of real-time information to share with orchestrators. There are tradeoffs which revolve around scalability and performance. More real-time data means better orchestrator computation results, fewer rounds of negotiation, and faster provisioning times. Too much real-time information can result in scalability issues and the need to ignore some updates in order to keep the control feedback loop stable. Future activities will seek to quantify and develop best practices regarding the amount, type, and update frequency of real-time data to be included in the resource model exchange.
- Real-time Data Dynamic Adjustment - The optimal amount and type of real-time information is expected to be dependent on specific deployment topologies, changing operational conditions, and service objectives. Future activities will investigate options of data collection, analysis, and mechanisms associated with the dynamic adjustment for real-time data inclusion in resource models.
- Resource Model Abstraction Level - Resource models can vary from highly detailed (complete representation of the physical infrastructure) to highly abstract (an entire infrastructure described as single node with just edge connections). Future activities will investigate and evaluate different levels of abstraction. There is a complex interaction between abstraction level and the real-time data issues noted above which will also be investigated.
- Policy Guided Decisions - There are multiple levels of authentication and authorization that are independently managed by different resource managers and orchestrators. Federated user authentication will allow individual resource owners to use a common identification base over which to apply their local policies. These policies can be communicated in the resource models and service computations. Future activities will investigate the best method for realizing multi-domain, multi-resource authentication and authorization, including issues around granularity and resource types (user, project, domain, individual network, end system resource elements, and/or flows).
- Machine Learning and Artificial Intelligence - The above issues highlight that SENSE is a real-time system with multiple levels of dynamic information exchange and feedback loops. Multiple architectural features are included to allow tailoring of this real-time information based on deployment, operational, and service constraints. Previous work has evaluated unsupervised. semi-supervised [45], and reinforcement learning [46] techniques to classify flows in real time and plan network usage profiles. Future activities will evaluate if these types of systems can leverage SENSE data as an added input and also recommend SENSE service provisioning actions. The system goals would include enhanced individual workflow quality of experience and optimization of overall infrastructure use.

As the SENSE architecture and implementation evolves through multi-institution testbed deployment, the focus of the project is to continue integration with domain science use cases and transition the SENSE services to production status for both the network and application operations.

## Acknowledgments

We appreciate the contributions to Intent APIs and Superfacility use case by Mariam Kiran and NERSC testbed deployment through Damian Hazen and Jason Lee. Work discussed in this paper was supported through multiple projects and grants from the Department of Energy and National Science Foundation including the following:

Caltech
- OLiMPS, DOE/ASCR, DOE award #DE-SC0007346
- SDN-Next Generation Integrated Architecture (SDN-NGenIA), DOE/ASCR, DE-SC0015527